\documentclass{vgtc}                          

\graphicspath{{figures/}{pictures/}{images/}{./}} 

\usepackage{times}                     

\usepackage{tabu}                      
\usepackage{booktabs}                  
\usepackage{lipsum}                    
\usepackage{mwe}                       

\usepackage{mathptmx}                  

\onlineid{2163}

\vgtccategory{Research}

\vgtcinsertpkg

\title{The Role of Mixed and Augmented Reality in Medical Visualization:\newline Literature Review and A Context-Aware Taxonomy}

\author{Xinrui Zou\thanks{e-mail: xzou8@jh.edu}\\ %
        \scriptsize Johns Hopkins University\\ %
        \scriptsize National Institutes of Health %
\and Mingxu Liu\\ %
     \scriptsize Johns Hopkins University %
\and Hongchao Shu\\ %
     \scriptsize Johns Hopkins University %
\and Ruixing Liang\\ %
     \scriptsize Johns Hopkins University %
\and Mathias Unberath\\ %
     \scriptsize Johns Hopkins University %
\and Alejandro Martin-Gomez\thanks{e-mail: am471@uark.edu}\\ %
     \parbox{1.4in}{\scriptsize \centering University of Arkansas}}

\teaser{
    \centering
    \includegraphics[width=0.99\textwidth]{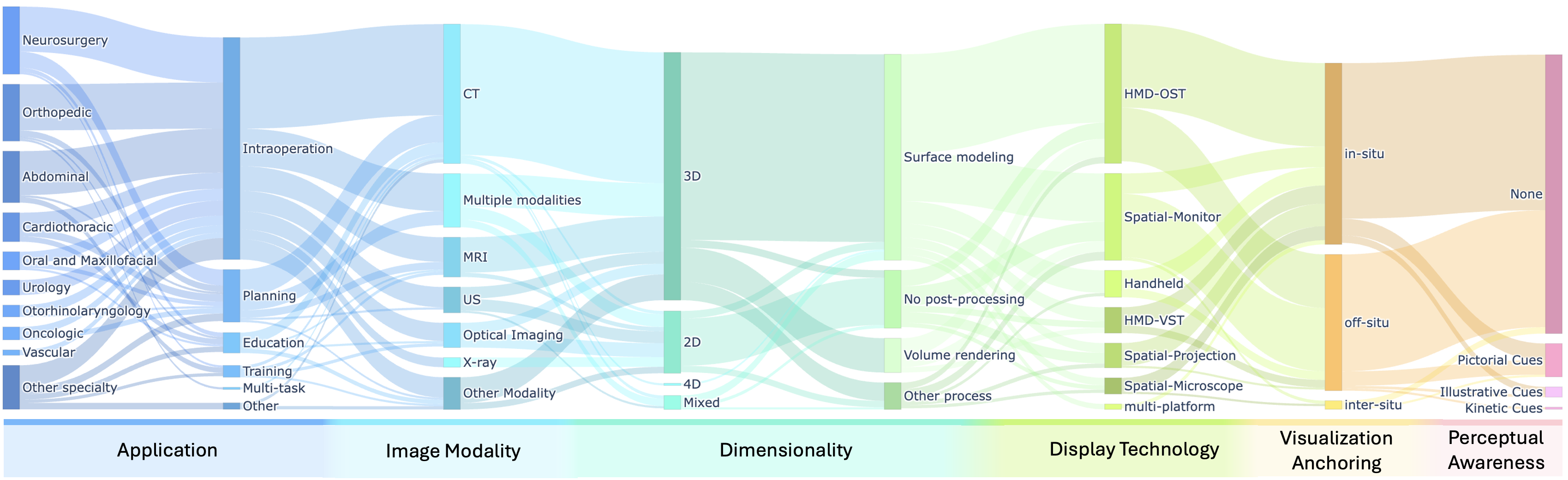} 
    \caption{Context-Aware Taxonomy of Medical Visualization in AR/MR. Our proposed taxonomy includes six distinctive categories: application (presented by medical specialty and clinical task), image modality, dimensionality (including rendering method), display technology, and levels of visualization anchoring and perceptual awareness. Node lengths in the Sankey diagram encode the relative prevalence of each sub-category, while flow width reflects co-occurrence patterns across studies.}
    \label{fig:sankey}
}

\abstract{
The discovery and evolution of medical imaging technologies have enabled non-invasive visualization of internal anatomy that has become essential for supporting diagnosis, monitoring, and treatment. However, because medical imaging relies on complex physical processes and contrast mechanisms for image formation, imaging alone is not sufficient to enable humans to leverage the resulting information fully. 
In addition, traditional methods to visualize the resulting information use two-dimensional displays to present three-dimensional anatomical structures.
The introduction of Augmented and Mixed Reality (AR/MR) technologies offers an opportunity to provide valuable paradigms for medical imaging visualization, allowing users to observe, explore, and interact with anatomical information in more spatially intuitive ways.
However, naive implementation without careful design considerations can lead to perceptual inconsistencies, potentially compromising utility and effectiveness.

In this paper, we present a structured taxonomy of medical AR/MR visualization strategies aimed at providing clearer insight into how visualization design varies across clinical use cases. The taxonomy organizes techniques based on four core design components: \textit{image modality}, \textit{data dimensionality}, \textit{display technology}, and \textit{clinical application}.
In addition, we introduce two critical dimensions that are often overlooked in the literature: \textit{visualization anchoring} (the spatial relationship between virtual content and the physical world), and \textit{perceptual awareness} (the use of visual cues to support spatial interpretation). Together, these components form a comprehensive taxonomy, offering a detailed framework for selecting appropriate visualization techniques in medical applications. 
} 


\begin{document}


\firstsection{Introduction}
\maketitle

Medical image visualization is the process of creating visual representations of anatomical structures from imaging data. These representations enable non-invasive visualization of internal anatomy and have become essential to support anatomical interpretation, clinical diagnosis, surgical planning, or image-guided surgical procedures.
Because medical imaging relies on complex physical processes and contrast mechanisms for image formation, imaging alone is insufficient to enable humans to fully leverage the resulting information. To make the information contained in medical images clinically relevant, effective techniques are needed to transform raw image data into meaningful visual representations. 

The use of visual representations to understand anatomy precedes the development of modern imaging technologies.
Leonardo da Vinci's 15th-century \cite{da1983leonardo} and Andreas Vesalius’s 16th-century illustrations
~\cite{Vesalius1555}, offered detailed insights into the human body using innovative drawing techniques~
that provided unprecedented views of internal anatomical structures. However, the creation of such illustrations was inherently invasive, as they would require an anatomical sample, and were limited to the representation of surface structures visible to the naked eye.
The discovery of X-rays by Röntgen in 1895 introduced a non-invasive method to observe internal anatomical structures, advancing medical diagnostics \cite{Roentgen1896}. In 1942, Dussik introduced a system capable of generating medical images using ultrasound signals, further expanding the field with radiation-free methods \cite{Dussik1942}. These technologies enabled physicians with unprecedented capabilities, providing a unique view of the human body. However, they remained constrained to the observation of a three-dimensional anatomy using two-dimensional images, complicating the interpretation of overlapping structures within complex tissues. The development of more advanced imaging technologies, like computed tomography (CT) in the 1970s and magnetic resonance imaging (MRI) in the 1980s, enabled the acquisition of volumetric datasets that captured internal anatomy in three dimensions, greatly enhancing spatial understanding. In the mid-1980s, these medical imaging modalities catalyzed the use of computer graphics methods to improve their visualization and inspired the development of techniques that could transform raw medical data into more intuitive representations. Algorithms like \textit{marching cubes} enabled efficient extraction of high-resolution isosurfaces directly from CT and MRI voxel grids \cite{10.1145/37401.37422}, and shortly thereafter, \textit{direct volume rendering} methods produced images by compositing voxel color and opacity along viewing rays \cite{levoy2002display,10.1145/54852.378484}.

Despite significant advancements in the development of imaging and processing technologies, the presentation of three-dimensional (3D) medical data has traditionally relied on two-dimensional (2D) media, such as physical films and monitors. Even with the advent of digital imaging systems in the mid-20th century, dynamic interaction with 3D data remained limited to 2D displays. As a result, clinicians had to mentally reconstruct spatial relationships, highlighting the need for more advanced display technologies capable of accurately presenting anatomical structures.

The introduction of immersive display technologies for the presentation of digital content in three dimensions, including virtual, augmented, and mixed reality (VR/AR/MR), offered a promising solution to this problem. VR facilitates the creation of immersive environments to interactively explore medical data, making it a viable alternative for educational settings. However, its separation from the real world limits its application in scenarios that require alignment with the patient’s anatomy, such as intraoperative procedures. \textit{AR and MR, by contrast, integrate virtual content directly onto the physical environment, enabling observation of spatially contextualized information and interactive content.} This capability allows clinicians to view and interact with patient-specific anatomy in real-time, making AR and MR suited for different clinical tasks. 
Although the term MR was originally conceived as an umbrella term to describe any system capable of enriching our perception of the real world by the integration of computer-generated content in the Reality-Virtuality Continuum \cite{milgram1994taxonomy}, including AR, it is now seen as a more robust form of AR, characterized by advanced environmental understanding that allows the virtual content to interact with the users and the surrounding environment \cite{speicher2019mixed}.

Early works showed the potential benefits of integrating AR/MR into medical practice in the 1990s \cite{bajura1992merging}. Since then, this technology has become more prevalent across surgical and diagnostic domains, where demand for precision, interactivity, and spatial awareness has increased in clinical decision-making. 
As AR/MR systems have matured, their design has become more complex, not only in terms of technical implementation but also in how visual content must be tailored to different clinical workflows and user needs.
This growing diversity has introduced new challenges in selecting appropriate visualization methods. Designers now must balance spatial accuracy, visual clarity, and usability, aligning the visualization strategy with the specific needs of a clinical task. Addressing these challenges requires more than just advances in rendering or display technology. It calls for a structured understanding of how the context can influence the visualization design space. 
The relationship between clinical context and selection of visualization strategies in medical AR/MR has not been systematically reviewed. Existing literature reviews focus on technical implementation or clinical use.
While practical studies discuss perceptual challenges associated with the selection, design, and presentation of augmented content in medical environments, these aspects remain comparatively underexplored, especially within the clinical context.

To bridge this gap, we conducted a systematic search and review, and introduced a context-aware taxonomy for medical image visualization in AR and MR  (\cref{fig:sankey}). This taxonomy considers the clinical context, emphasizing how visualization strategies are designed by task demands. It categorizes visualization techniques not only by image modality, data dimensionality, display technology, or clinical use case, but also highlights two essential and often overlooked dimensions: \textit{visualization anchoring} (the spatial relationship between virtual content and the real world) and \textit{perceptual awareness} (the use of visual cues to support spatial interpretation). These dimensions are critical for supporting safe and effective use of AR/MR in clinical practice. 
This context-aware taxonomy aims to provide a reference framework for researchers and developers to support the design, evaluation, and integration of AR/MR visualization techniques into clinical workflows\footnote{Results from this review, including interactive tables and resources, are publicly available here: https://xzjiu.github.io/Mixed-and-Augmented-Reality-in-Medical-Visualization/articles.html}.

\section{Related Work}
The widespread integration of AR and MR in medical applications over the past years has motivated the creation of review works that classified this technology from usage (e.g., clinical tasks) or design perspectives (e.g., general pipelines). However, a limited number of studies have systematically examined how clinical objectives influence the design of visualization strategies, particularly in terms of spatial integration and perceptual support.

A large body of review literature emphasizes the clinical applications of AR/MR across medical specialties, such as orthopedics \cite{mccloskey2023virtual,hersh2021augmented} and plastic surgery \cite{kim2017virtual}, as well as task-specific areas like education \cite{tang2020augmented,adapa2020augmented} and surgical guidance \cite{kersten2013state}. While these reviews offer valuable insights into system performance, clinical feasibility, and workflow integration, they typically prioritize procedural outcomes over visualization design. As a result, there is still a limited understanding of how visualization designs are systematically adapted to different clinical scenarios.
Other studies have systematically categorized visualization approaches based on their visual properties. 
Existing frameworks describe important parts of the design space, such as the progression from raw data to rendered display in image-guided surgery \cite{kersten2011dvv}, visual properties such as compositing style, visibility, and interactivity \cite{zollmann2020visualization}, or hardware and display configurations such as optical- versus video see-through displays and egocentric versus exocentric views \cite{bimber2005spatial, sielhorst2008advanced, normand2012new, ma2023visualization}. These frameworks provide useful foundations, but they emphasize different technical layers of AR/MR visualization. They do not explicitly connect these layers to medical tasks, image modality, anatomical constraints, anchoring strategy, and perceptual support. As a result, they offer limited guidance for understanding how visualization choices vary across clinical contexts.

The lack of a structured framework linking visualization design to clinical context complicates analyzing how specific medical tasks, data types, or workflow constraints influence design choices in AR/MR systems. 
To address this gap, our taxonomy links technical dimensions such as image modality, dimensionality, and display technology to medical use cases, while highlighting the importance of visualization anchoring and perceptual awareness in accounting for content placement and spatial interpretation.

\section{Data Collection}
\begin{figure}
        \centering
        \includegraphics[width=0.33\textwidth]{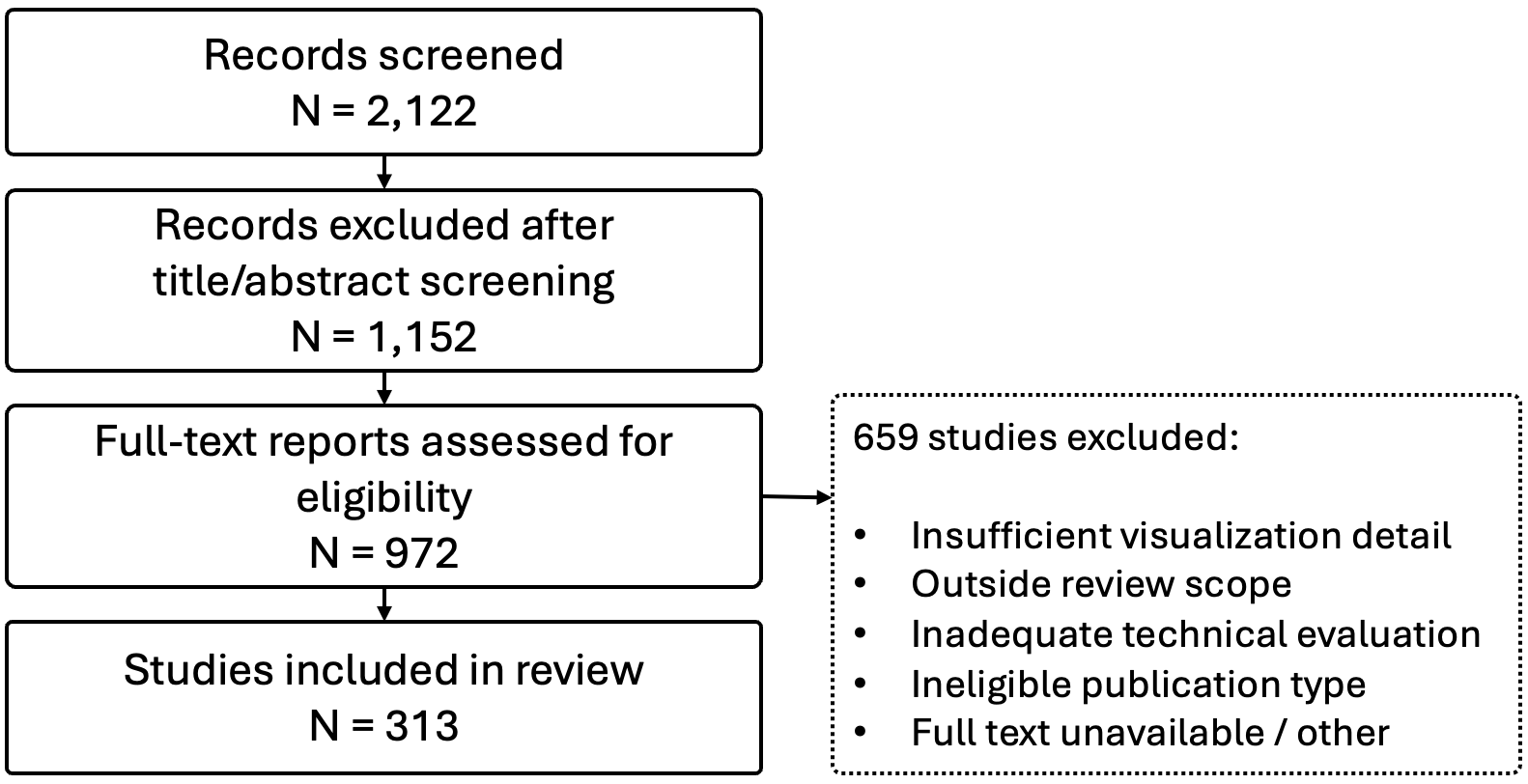} 
        \caption{Literature screening and eligibility assessment process.}
        \label{fig:prisma}
\end{figure}

To examine how AR/MR have supported imaging visualization in clinical settings, we conducted a structured literature review based on a systematic search of peer-reviewed studies in major medical and technical databases. Given their coverage of medical, engineering, and technology-oriented research, we used PubMed, IEEE Xplore, and Scopus as primary sources to identify studies explicitly describing the representation, rendering, or integration of medical imaging data using AR/MR in clinical applications. Search and selection reporting was informed by applicable PRISMA 2020 items \cite{page2021prisma}. Database-specific search strings and filters are provided in the supplementary material.

\begin{itemize}
    \item \textbf{PubMed}: [\textit{Augmented Reality} OR \textit{Mixed Reality}] AND \newline [\textit{Visualization}] in [\underline{Title} OR \underline{Abstract}]
    \item \textbf{IEEE Xplore}: [\textit{Augmented Reality} OR \textit{Mixed Reality}] AND [\textit{Visualization}] in [\underline{Title} OR \underline{Abstract}] AND [\textit{Medical} OR \textit{Surgery}] in [\underline{Title} OR \underline{Abstract}]
    \item \textbf{Scopus}: [\textit{Augmented Reality} OR \textit{Mixed Reality}] AND [\textit{Visualization}] AND [\textit{Medical} OR \textit{Surgery}] in [\underline{Title} OR \underline{Abstract}]
\end{itemize}

The literature search covered studies published before February 5, 2026. Built-in database filters were used, where available, to exclude reviews and meta-analyses and to retain peer-reviewed original articles. 
After duplicate removal and dataset cleaning, two reviewers independently screened titles and abstracts, followed by full-text eligibility assessment.
During screening, we excluded studies unrelated to AR/MR-based medical image visualization, including papers outside medical contexts and those focused solely on hardware, registration, interface design, or navigation without relevant medical image visualization. Full-text review further evaluated technical relevance, excluding reports that lacked sufficient detail on image representation in the AR/MR system, fell outside the scope of medical image visualization, provided inadequate technical or evaluative information, or met ineligible publication criteria (Fig.~\ref{fig:prisma}). A total of 313 studies were included in the final review. 

Information from selected studies was recorded using a shared template that defined the taxonomy dimensions and decision rules used to classify them. At the full-text stage, four authors contributed to the extraction, and each study was reviewed by two authors. Disagreements or uncertain classifications were resolved through discussion. For each study, categories were assigned based on the system description, figures, and implementation details reported in the paper.

\section{Taxonomy}
The taxonomy presented in this work provides a structured framework for analyzing and designing AR/MR visualization systems in medical contexts. By organizing key aspects of such visualization into six dimensions, this taxonomy serves as a guide for analyzing and selecting techniques that meet the unique challenges of medical applications. This section presents an overview of the taxonomy, with subsequent sections providing more detailed information regarding every individual dimension.\vspace{3pt}

\noindent \textbf{\underline{Application:}}~The clinical task and medical specialty. Differences in application affect overall visualization requirements.\vspace{3pt}

\noindent \textbf{\underline{Image Modality:}}~The medical imaging data used by the system. This review adopts a broad definition of medical imaging, including traditional radiological modalities (e.g., CT, MRI, Ultrasound) and optical imaging aimed at visualizing anatomical structures for clinical or educational use \cite{ganguly2010medical}.\vspace{3pt}

\noindent \textbf{\underline{Dimensionality:}}~The spatial structure of the imaging data, regardless of its visual presentation or perceptual interpretation. Most AR applications present information as 2D slices or 3D models. Time-resolved imaging data can also be visualized as 4D content (e.g., cardiac sequences). \vspace{3pt}

\noindent \textbf{\underline{Display Technology:}}~The hardware used to deliver the augmented content, selected based on image modality, visualization format, and interaction needs. These technologies include head-mounted displays (HMDs), hand-held devices (e.g., smartphones), or spatially-located systems (e.g., monitors). \vspace{3pt}

These four core dimensions form the foundation of any medical AR/MR application. They define what data is used, how it is visualized, where it is displayed, and in what clinical context. However, a fully optimized system requires additional considerations to ensure effective integration into real-world medical environments. To improve usability, perceived spatial accuracy, and clinical relevance, these systems should include two additional dimensions: \vspace{5pt}

\begin{figure}[]
        \centering
        \includegraphics[width=0.5\textwidth]{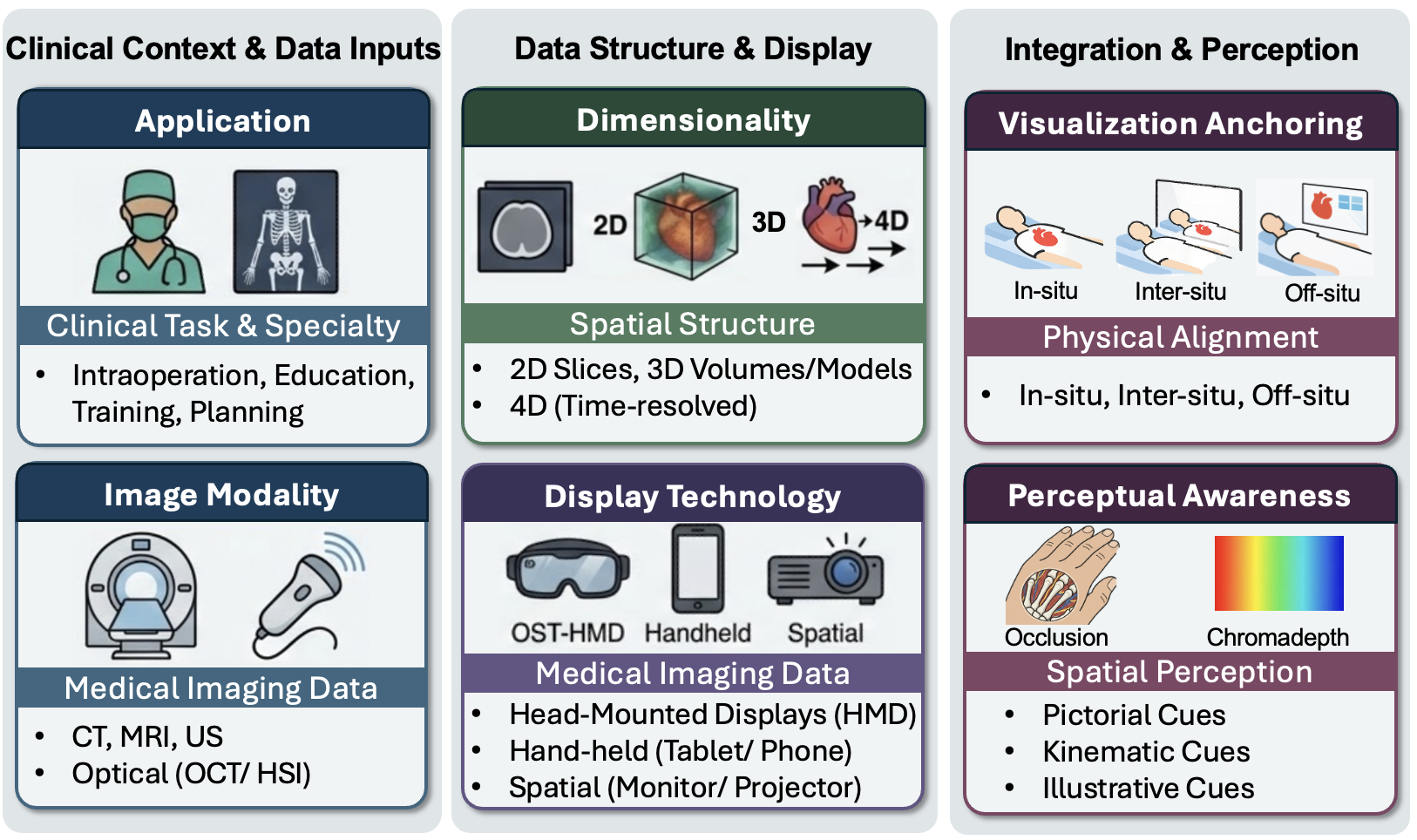} 
        \caption{Taxonomy of medical AR/MR visualization organized by design considerations across \textit{application}, \textit{image modality}, \textit{dimensionality}, \textit{display technology}, \textit{visualization anchoring}, and \textit{perceptual awareness}.}
        \label{fig:six_dimension}
\end{figure}

\noindent \textbf{\underline{Visualization Anchoring:}}~Determines how augmented content is presented in the physical environment. Virtual content can be overlaid on the patient (in-situ), positioned adjacent to the anatomy (inter-situ), or displayed separately (off-situ). Unlike registration, which refers to the geometric mapping between image-derived content and the patient or scene, anchoring describes the resulting placement of visual content in the user’s view.\vspace{3pt}

\noindent \textbf{\underline{Perceptual Awareness:}}~Influences how users perceive virtual objects within the physical environment, playing a significant role in depth perception, spatial alignment, scene understanding, and visual consistency. The use of techniques that integrate fundamental aspects of visual perception in their design can contribute to enhanced accuracy during task performance and increase system acceptance in complex procedures. \vspace{5pt}

This taxonomy offers a structured perspective on the medical AR/MR systems across six dimensions. While the following sections introduce each individually, the \textit{Application} (\cref{sec:application}) is presented last to synthesize the findings from a clinical workflow perspective and provide a context-aware overview of the field.

\section{Image Modality}

The clinical context plays a determinant role and determines the imaging modality presented using AR/MR. Rigid bony anatomy is commonly visualized using CT or X-ray imaging technologies, while soft-tissue anatomy often requires the use of MRI. When anatomy deforms or moves, ultrasound imaging can provide intraoperative updates, while multimodal combinations can help combine and visualize rigid and deformable structures \cite{wong2023mixed,gerard2018combining,chen2023towards}.
\vspace{3pt}

\noindent \textbf{\underline{Radiography (X-rays):}}~Used in procedures that require intraoperative visualization of bony anatomy, such as fluoroscopy-guided interventions, provides a two-dimensional projection of the patient's anatomy. The use of AR/MR in these settings can contribute to reduced mental effort by placing the two-dimensional images in the three-dimensional spatial context \cite{moglia2023mixed}, improving scene understanding and coordination during guidance \cite{fotouhi2019interactive,pauly2015machine}. Reported benefits also include fewer repeated acquisitions in some settings, with potential implications for radiation exposure \cite{weidert2019video}. The main constraint is that content remains projection-based, so depth inference and occlusion reasoning depend heavily on visualization and interaction design rather than image data alone.\vspace{3pt}

\noindent \textbf{\underline{Computed Tomography (CT):}}~The most prevalent modality in medical AR/MR. It allows for generating surface models and volumetric representations of the anatomy \cite{shekhar2010live,bichlmeier2007laparoscopic,carl2019augmented,de2019augmented}. In most cases, CT is acquired preoperatively and used for planning, guidance, and education \cite{reitinger2004tools,olexa2024apple,zhang2023feasibility,rose2019development,ma2016personalized,samuel2024visuo}. 
CT may be used in interventions where point-of-care acquisition is feasible \cite{rieder2024augmented}. Intraoperatively, CT can provide on-demand updates but remains limited by cost, radiation exposure, and operating room constraints \cite{umebayashi2018augmented,carl2019augmented}.\vspace{3pt}

\noindent \textbf{\underline{Magnetic Resonance Imaging (MRI):}}~This imaging modality is used when soft-tissue visibility is crucial, with common applications in neurosurgery \cite{suetens2017fundamentals,lawonn2018survey}. In AR/MR, MRI is typically acquired preoperatively, but intensity variability and comparatively lower resolution and signal-to-noise ratio often necessitate preprocessing or segmentation for reliable visualization. Intraoperative MRI has been incorporated into AR workflows to update anatomy and reduce mismatch from brain shift or resection \cite{fritz2014mr}. However, the intraoperative use of this technology is less frequent due to cost, acquisition time, and infrastructure requirements.\vspace{3pt}

\noindent \textbf{\underline{Ultrasound:}}~This modality is used when real-time visualization is required, especially for intraoperative guidance where anatomy deforms or moves. Most AR/VR systems provide 2D ultrasound overlays to preserve spatial context \cite{costa2023augmented}, while some reconstruct 3D volumes for visualization \cite{von2023augmenting}. The core limitation of this modality is its image quality and associated interpretability, which motivates visual enhancement techniques such as shading to improve readability \cite{maddali2024monte}. Although 4D ultrasound is common in fetal and cardiac imaging, it remains rare in AR/MR, likely due to the computational and visualization demands of time-varying volumes.\vspace{3pt}

\noindent \textbf{\underline{Optical Imaging:}}~Optical imaging appears most often as endoscopic video in minimally invasive procedures. In many AR/MR systems, it serves as a contextual background layer for overlays and is, therefore, not treated as a standalone modality in our taxonomy. Other systems use this modality to reconstruct and augment the operative scene \cite{gonzalez2023anatomic}. Alternative optical modalities are less common but illustrate emerging directions, including hyperspectral imaging for tissue classification \cite{sancho2023slimbrain,huang2020augmented} and optical coherence tomography for microstructural cross-sectional visualization \cite{roodaki2015introducing}.

\section{Dimensionality}
Dimensionality refers to the spatial and temporal dimensions of the imaging data and the representation strategies used to visualize them in AR/MR. Because dimensionality constrains what can be rendered and registered, it shapes downstream choices such as slice- or projection-based displays, segmented surface models, and volume rendering. Although some systems use 4D imaging to represent time-varying anatomy, such as respiratory motion or cardiac cycles \cite{bindschadler2022heartbeat4d}, broader adoption is limited by acquisition burden, real-time processing demands, and workflow integration overhead. Accordingly, this section focuses on the prevailing 2D and 3D representations and the processing choices most often paired with them.\vspace{2pt}

\noindent \textbf{\underline{Two-dimensional:}}~These images are used in AR/MR systems in two primary ways. First, they can serve as direct anatomical references without extensive processing. Examples include X-rays \cite{fotouhi2020development}, 2D ultrasound \cite{kang2014stereoscopic}, or CT/MRI slices \cite{fritz2014mr}, which may be overlaid onto the patient’s body or positioned in space to support intraoperative guidance.
Second, 2D content can be displayed on virtual panels within the user’s field of view, functioning like movable screens. These panels typically combine medical images with contextual information such as vital signs, procedural steps, or annotations, offering an easily accessible reference without requiring anatomical registration. Such displays can improve ergonomics and cognitive flow by allowing flexible positioning within the surgical workspace \cite{stewart2022study}. Although 2D images themselves lack depth information, they remain a practical component of medical AR/MR systems.\vspace{2pt}

\noindent \textbf{\underline{Three-dimensional:}}~In medical AR/MR, 3D data is typically post-processed using one of two approaches: volume rendering or surface modeling. Volume rendering displays full volumetric datasets by mapping voxel intensity values to color and transparency, allowing for direct visualization of internal and external anatomical features, but it is computationally expensive.
Compared to direct volume rendering, surface modeling reduces processing complexity by converting segmented anatomical regions into lightweight and interactive mesh-based models, making it the most commonly used form of 3D visualization in AR/MR systems.
However, this method is highly dependent on the quality of segmentation. Inaccurate or incomplete segmentation can result in misleading visualizations \cite{sielhorst2008advanced}. 
Regardless of processing method, 3D visualization enables a level of spatial interaction not achievable with 2D formats, which has made it the predominant form of visualization in medical AR/MR. Users can inspect internal features and explore complex spatial relationships, capabilities essential for tasks that require detailed anatomical understanding. As a result, 3D representations are widely employed in AR/MR systems for surgical planning \cite{schwenderling2022augmented}, procedural training \cite{viglialoro2018augmented}, and medical education \cite{da2017segmented}.

\section{Display Technology}


Display technology shapes not only how augmented information is perceived, but also how it fits into clinical workflows. In procedural environments, practical constraints such as sterility and hands-free operation, line-of-sight to the operative field, team-shared viewing, and tolerance to latency often dominate device selection. More importantly, display choice is tightly coupled with visualization strategy as it influences whether information can be placed directly in the surgical field, how occlusion and depth consistency can be handled, and how much perceptual support is needed to maintain interpretation. In this work, we group displays by their usage context into head-mounted, hand-held, and spatial devices, and summarize their characteristic workflow trade-offs.\vspace{3pt}

\noindent \textbf{\underline{Head Mounted Displays:}}~
Present content from a user-centered viewpoint and support hands-free use, which is attractive for spatial tasks in the operating room \cite{condino2019perceptual}. Optical see-through head-mounted displays (OST-HMDs) preserve direct vision of the surgical field, even during overlay loss or system failure, but are limited by restricted field of view, reduced contrast, calibration sensitivity, and close-range visual discomfort \cite{grubert2017survey,gsaxner2021augmented}. They also cannot reliably occlude real objects, compromising depth consistency when overlays are intended to appear in front \cite{hong2011three}. Video see-through head-mounted displays (VST-HMDs) composite the real and virtual content, enabling controlled occlusion and more consistent depth composition \cite{edwards1993video,cutolo2018perspective}. Their main drawbacks are latency and dependence on the camera and video-processing pipeline. Interruptions in capture, tracking, or rendering can cause lag or loss of the real-scene view, which can disrupt hand-eye coordination \cite{birlo2022utility,guo2019online}.

\noindent \textbf{\underline{Hand Held Devices:}}~
Hand-held devices are used when portability and low deployment cost are prioritized, such as point-of-care viewing and education \cite{zhou2008trends,kenngott2018mobile,gurses2024interactive}. Their main limitation is that they require holding and manual interaction, conflicting with hands-free and sterile requirements in many intraoperative workflows. \vspace{3pt}

\noindent \textbf{\underline{Spatial Displays:}}~
Spatial displays use fixed displays that can be viewed by the clinical team. Within this category, \emph{monitors} are the most common display platform because they fit naturally into endoscopy, fluoroscopy, and robotic surgery workflows with minimal additional hardware \cite{bichlmeier2007laparoscopic,pauly2015machine}. Their availability in the operating room facilitates their integration into existing settings. However, the spatial mismatch between their location and the patient's body can increase cognitive load in spatial tasks \cite{li2023comparative}. When additional depth cues are desired, some systems employ \emph{stereo} or \emph{autostereoscopic} monitors, but their use remains limited by practical constraints such as eye-wear requirements, restricted viewing zones, and resolution trade-offs \cite{kang2014stereoscopic,kang2022autostereoscopic,huang2021augmented,wang2017autostereoscopic}. Surgical microscopes offer high optical fidelity and a familiar view-aligned platform for overlaying guidance in microscopy procedures \cite{kersten2015augmented,roodaki2015introducing}, but their fixed setup restricts mobility and interaction. Projection-based systems enable shared visualization of overlays without wearing a device. They can reduce gaze shifting, but they are sensitive to viewpoint changes and surface geometry. These changes can introduce parallax and distortion without robust calibration and compensation \cite{bimber2005spatial,gao2021projector,baumeister2017cognitive,wu2022stereotactic,wen2013projection}. Beam splitter setups similarly support direct view while injecting overlays into the line-of-sight, but their effective viewing range can degrade with angle changes \cite{sielhorst2008advanced}.

\section{Visualization Anchoring}

Visualization anchoring describes how imaging-derived content is positioned relative to the patient and the operative scene. Across medical AR/MR systems, augmented information can be displayed on a separate screen away from the physical site, placed near the patient while remaining spatially separated from the interaction region, or overlayed directly onto the anatomy. We classify this relationship using three anchoring classes, off-situ, inter-situ, and in-situ, which reflect increasing degrees of spatial coupling between virtual content and the physical target. 
This anchoring choice is not only a presentation detail. It mediates attention switching, hand–eye coordination demands, and the extent to which accurate registration and perceptual cues are needed.

\subsection{Off-situ}
Off-situ visualization refers to the display of medical content away from its physical source with virtual elements that are not spatially registered to the patient's anatomy. Instead, they appear in separate locations, requiring users to shift attention and mentally map the information back to the patient.
An example of this visualization is the use of external monitors, where augmented content (e.g., navigational information) is viewed separately from the surgical field. This setup maintains workflow compatibility but increases cognitive load due to the need for mental alignment.
Off-situ visualization can also be observed even when using HMDs. The presentation of virtual content that is not spatially registered to the real world, as in the case of floating 3D models \cite{fick2023comparing} or virtual display panels \cite{stewart2022study} displayed in a fixed screen space, would be classified as off-situ.
Off-situ visualization is particularly prevalent in medical education, where approximately 78\% of the reviewed literature adopted this approach. It allows users to freely interact with 3D anatomical models, explore spatial relationships, and engage with complex structures without requiring data registration. This enables safe and repeatable training conditions and effective conceptual learning.

\subsection{Inter-situ}
Inter-situ visualization is observed when the augmented content is presented outside the immediate physical space while maintaining a visual reference to real-world objects \cite{yu2024advanced}. Examples of these methods include mirror-based approaches that deliver the AR experience using real mirrors \cite{martin2020augmented} placed in the environment or duplicated AR, which replicates regions of interest in a separate virtual space for flexible interaction \cite{yu2022duplicated}.
A key advantage of inter-situ visualization is its ability to reduce visual obstruction while preserving spatial alignment. Existing studies have shown that separately positioned but rotationally aligned models contribute to increased user preference, reducing interference with the region of interest while maintaining a clear spatial relationship with real-world structures \cite{benmahdjoub2023evaluation}.
In addition, inter-situ visualization supports multi-user collaboration without physical interference, facilitates seamless real-virtual transitions, and enables flexible content modifications such as clipping or scaling \cite{yu2022duplicated}.

\subsection{In-situ}
In-situ visualization presents virtual content directly over the physical anatomy or object of interest, enabling real-time spatial alignment between medical data and the patient’s body~\cite{bichlmeier2007contextual}. This spatial coherence supports in-situ visualization anchoring and reduces the hand–eye coordination demands commonly seen with off-situ displays, where users must mentally map between separated visual and physical spaces \cite{nguyen2022holous}. Both 2D~\cite{nguyen2022holous, marker20171} and 3D content \cite{moreta2021combining} can be integrated using in-situ techniques. In medical applications, in-situ visualization is used to support tasks such as preoperative planning \cite{moreta2021combining} and intraoperative guidance \cite{marker20171}.
In our review, more than 50\% of the surveyed systems employed in-situ visualization, highlighting its prevalence in clinical applications. However, the effectiveness of this visualization often relies on accurate registration between the virtual and real world. Misalignment may cause incorrect spatial interpretation, which is critical in delicate procedures. Another challenge is the consistent integration of visual cues like occlusion and depth. Without proper handling, virtual content appears detached from the body, reducing clarity and reliability.

\section{Perceptual Awareness}
\label{sec:perceptual}
A central challenge in AR-based medical visualization lies not only in spatially registering virtual content to the clinical environment, but in ensuring that users can accurately interpret the spatial relationship between virtual and real objects. Even with precise alignment, poor visual integration, such as missing occlusion or weak depth cues, can result in misjudging anatomical structures or tool placement, potentially compromising clinical tasks \cite{sielhorst2006depth}.
Integrating visual cues can help users perceive virtual content more accurately within the physical scene.
While some systems (e.g., stereoscopic displays) enhance the sense of virtual depth, they do not ensure awareness of spatial relationships between virtual content and the physical world.
Achieving effective perception in AR/MR systems requires the thoughtful design of visual cues that integrate virtual elements seamlessly into the physical scene.

    \subsection{Pictorial Cues}
    These cues allow us to extract information from 2D images to perceive depth and estimate three-dimensional spatial relationships between the objects in a scene using a single eye. Unlike hardware-dependent cues such as stereopsis, these cues can be applied across a wide range of AR/MR platforms, regardless of the display technology used to present the augmented content.

        \subsubsection{Occlusion}
        Occlusion is the most powerful pictorial cue. When one object partially blocks another, we naturally perceive it as being in front. In medical AR/MR, this principle is critical for correctly conveying the location of anatomical structures that are not directly visible. If not handled properly, virtual content may appear outside the body even when correctly registered, leading to perceptual errors \cite{kalia2016interactive}. This problem is often observed when anatomical structures located inside the patient's body are naively rendered on top of their skin, causing the content to be perceived outside. Techniques that have been developed to address this issue include: \vspace{2pt}

        \noindent \underline{Virtual Cutaways:} Virtual cutaway techniques selectively remove or modify real-world surfaces to reveal underlying virtual structures while maintaining spatial context. In these techniques, a virtual window simulating an opening in the anatomy allows visualization of deeper structures while preserving the external surface \cite{bajura1992merging}. Similar approaches apply structured textures, such as random-dot masks, to the surface to enhance depth perception and improve the illusion of anatomical openings \cite{otsuki2015analysis}. These methods provide an intuitive way to explore hidden anatomy without disrupting spatial awareness. However, both techniques have limitations: virtual windows may obscure surface features or restrict visibility when applied to complex anatomy, while random-dot masks can introduce visual clutter or hide parts of the virtual structure, despite preserving spatial relationships \cite{wang2017autostereoscopic}.
        \vspace{2pt}
        
        \noindent \underline{Structured Occlusion:} Structured occlusion techniques aim to maintain spatial depth perception without fully removing the obstructing surface. Contour-based rendering techniques enhance the visibility of occluded structures by outlining their edges with adaptive contours, preserving depth relationships, and keeping the primary surgical field unobstructed \cite{kalkofen2007interactive}. Similarly, semi-transparent overlays can be used to blend real and virtual content,  partially revealing internal structures while retaining the surface texture of the occluding anatomy, reducing visual ambiguity \cite{bichlmeier2007contextual}.\vspace{2pt}
        
        \noindent \underline{Foreground Segmentation:} These methods use real-time depth sensing or machine learning-based classification to dynamically distinguish between physical and virtual elements. By accurately segmenting the surgeon’s hands, tools, or patient’s skin, this techniques ensure that foreground objects correctly occlude augmented content when necessary. Such techniques have been applied in contexts such as AR-assisted fluoroscopic guidance \cite{pauly2015machine, heiliger2023phantom}.

        \subsubsection{Texture}
        Texture density is a well-established pictorial depth cue, where surfaces appear less textured as they recede into the distance \cite{8115410}. Although these cues are generally more useful at larger distances, their fundamental principle has been adapted and integrated into medical visualization using texture gradients that can convey spatial information. In vascular visualization, procedural textures with stroke-width variations can represent depth, where thicker strokes indicate closer structures and thinner strokes depict deeper vessels \cite{4015442}. Although studies have shown that texture alone has minimal impact on depth perception compared to other cues \cite{8115410}, and an evaluation of texture density in AR concluded that it did not significantly enhance depth perception \cite{10670456}, this method can be integrated with chromadepth encoding to enhance spatial perception \cite{8115410}. 

        \subsubsection{Illumination}
        In our taxonomy, illumination-based effects such as shadows and reflections are grouped under pictorial cues, as they contribute to depth and shape perception through image-based visual information. While not traditionally considered a pictorial cue, reflection can provide valuable visual information about the spatial structure of a scene and reveal details about hidden objects. Therefore, we include it here under a broader interpretation of pictorial cues. \vspace{2pt}
        
        \noindent \underline{Shadow:}
        Shadows can be used to estimate the relative position and distance between objects by projecting areas of reduced illumination onto surfaces. In medical AR/MR, projecting shadows onto a reference plane can enhance spatial understanding by conveying how virtual elements relate to the physical environment \cite{wang2016inverse}. \vspace{2pt}

        \noindent \underline{Reflection:}
        Reflections, such as those provided by physical/virtual mirrors or polished materials, can provide additional viewpoints by displaying virtual objects over their surfaces. This effect reinforces spatial relationships by showing objects from multiple angles. Augmented mirrors, for example, have been explored in laparoscopic surgery to enhance depth perception by providing an alternative view of anatomical overlays \cite{bichlmeier2007laparoscopic, martin2020augmented}.  

    \subsection{Kinetic Cues}
    Kinetic depth cues convey spatial information through motion-based perception, where relative movement between the observer, the scene, or objects within the environment enhances depth understanding. Unlike pictorial cues, kinetic cues dynamically provide additional information as the observer’s perspective changes.
    \vspace{4pt}
    
    \noindent \textbf{\underline{Focus Modulation:}}
    Observer-driven techniques dynamically adjust visual parameters (e.g., transparency or clarity) to emphasize key regions \cite{bichlmeier2007contextual}. Modulating video image transparency based on the patient’s skin topology and the observer’s viewpoint can improve spatial perception. As the observer moves, the focus area updates, simulating an interactive scanning effect for internal structures.\vspace{3pt}

    \noindent \textbf{\underline{Dynamic Depth Representations:}}
    These techniques use real-time motion to encode depth, offering an alternative to static cues. Examples include surface waves radiating across objects and motion along anatomical pathways, both leveraging temporal variation to convey distance and shape \cite{martin2023closer}.
    These methods allow users to infer shape and structure via temporal depth variations, complementing static pictorial cues with movement-induced depth perception.

    \subsection{Illustrative Cues}
    Illustrative cues refer to explicitly designed visual encoding that communicate depth or spatial relationships through artificial means, such as color mapping or numeric annotations. Unlike pictorial cues, these require intentional design choices and may depend on user familiarity with the encoding scheme.\vspace{5pt}
    
    \noindent \textbf{\underline{Chromadepth:}}
    A method that maps the depth of an object to a specific color, typically assigning warmer colors (e.g., red) to closer objects and colder colors (e.g., blue) to distant ones. It enhances spatial awareness by creating a depth gradient that improves differentiation of anatomical layers and procedural landmarks \cite{kersten2013evaluation}.\vspace{3pt}
    
    \noindent \textbf{\underline{Direct Measurement:}}
    These cues provide explicit numerical depth information, offering precise spatial awareness. These cues include dimension lines, numeric distance annotations, and spatial markers that assist users in understanding relative positioning and scale. Examples of this approach include the use of real-time distance indicators between surgical tools and anatomical targets to support intraoperative guidance and improve procedural accuracy \cite{choi2016effective}.

\begingroup
\setlength{\tabcolsep}{4pt} 
\begin{table*}[ht]
\renewcommand{\arraystretch}{1.3}
\centering
\scriptsize
\caption{Summary of anatomical and procedural characteristics across medical specialties relevant to AR/MR system design.}
\begin{tabular}{p{2.2cm} p{2.3cm} p{3.8cm} p{2.2cm} p{6.0cm}}
\toprule
\multicolumn{1}{c}{\textbf{Specialty (N)}} & \multicolumn{1}{c}{\textbf{Structural Rigidity}} & \multicolumn{1}{c}{\textbf{Motion Complexity}} & \multicolumn{1}{c}{\textbf{Imaging Modality}} & \multicolumn{1}{c}{\textbf{AR/MR Design Considerations}} \\
\midrule
Neurosurgery (66) & Semi-rigid & Brain shift & MRI & Requires intraoperative compensation \\
Orthopedic (55) & Rigid & Minimal & CT/X-ray & Enables highly accurate static overlays \\
Abdominal (51) & Deformable & Organ motion, respiration & CT/MRI/US & Real-time update or deformable model support \\
Cardiothoracic (30) & Deformable & Cardiac, respiratory motion & CT/US & Constant updates and motion compensation needed \\
Maxillofacial (22) & Rigid+Deformable & Minimal & CBCT/Surface & Accurate for bone. Less reliable for soft tissue \\
Urology (17) & Deformable & Displacement, posture dependent & MRI/US & Limited landmarks. Intraop imaging often required \\
\bottomrule
\end{tabular}
\label{tab:medical_specialty}
\end{table*}
\endgroup

    \begin{figure*}[t]
        \centering
        \includegraphics[width=0.975\linewidth]{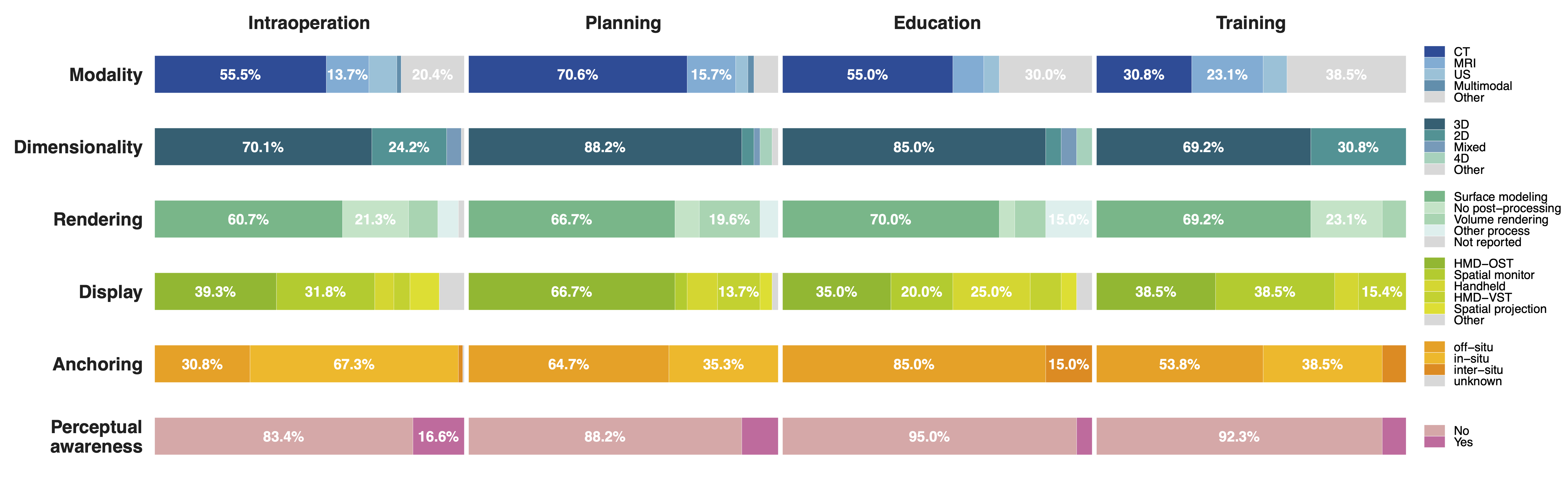} 
        \caption{Taxonomy patterns of visualization design choices across clinical tasks. Each row corresponds to one taxonomy dimension (modality role, dimensionality, rendering, display technology, anchoring, and perceptual awareness), and each column summarizes within-task proportions for a clinical task category (with sample size $n$ shown in the header). Percentage labels indicate the dominant categories within each task.}
        \label{fig:task_taxonomy}
    \end{figure*}

        \begin{figure*}[t]
        \centering
        \includegraphics[width=0.975\linewidth]{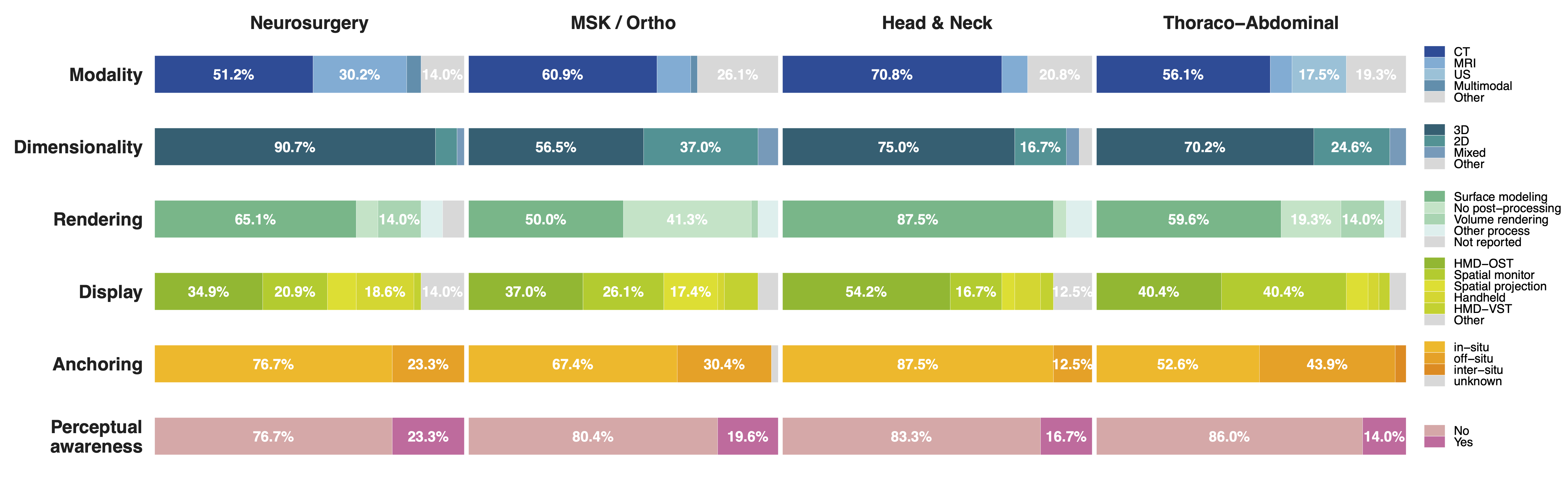} 
        \caption{Taxonomy profiles of intraoperative AR/MR visualization design choices stratified by surgical area. Each row corresponds to one taxonomy dimension, and each column summarizes within-area proportions. Percentage labels indicate dominant categories.}
        \label{fig:area_taxonomy}
    \end{figure*}

\section{Application}
\label{sec:application}
To organize the studies by application, we use two axes: the \textit{clinical task} within the workflow and the \textit{medical specialty}. We focus on clinical tasks first, because they determine how AR/MR is used and what visualization approaches are followed. We then use medical specialty to further stratify intraoperative systems, as these constrain the imaging modalities used, affecting registration and image readability.
We organized the included studies into four task categories along the clinical workflow: medical and patient education, simulation-based training, preoperative planning, and intraoperative use. Because category sizes are imbalanced, we report within-category proportions rather than absolute counts. Of the 313 included studies, 66.5\% address intraoperative applications, followed by preoperative planning (15.7\%), medical and patient education (6.4\%), and simulation-based training (4.2\%). Sparsely represented or non-workflow studies, such as standalone technical evaluations, are summarized in \cref{fig:sankey} but excluded from detailed comparison. The subsections proceed from education to intraoperative use, with intraoperative systems further examined by surgical area.

\subsection{Medical and Patient Education}
Educational applications use AR/MR to facilitate understanding of anatomical structures and medical procedures for students and patients. These systems can support intuitive exploration without requiring real-time imaging or precise anatomical alignment, and they are often based on pre-generated models derived from CT or MRI. Many tools prioritize conceptual understanding over spatial precision by providing interactive 3D models that can be inspected from different viewpoints and augmented with colors or annotations to aid comprehension \cite{gurses2024interactive,da2017segmented,ma2016personalized}. Perceptual support techniques are rarely reported explicitly in this category, likely because educational systems generally do not require tight geometric alignment or depth-critical \emph{in situ} interpretation. Instead, support for understanding is more often embedded in simplified presentation and interaction design than framed as a dedicated perceptual cueing strategy.

The relatively small number of educational AR/MR systems may reflect the fact that many educational objectives are commonly addressed using VR, which provides a controlled visual environment. However, the distinctive value of AR/MR in education lies not merely in placing virtual models in the real world, but in grounding learning in a physical context. For example, AR/MR can relate anatomy to a real body, manikin, or clinical tools, and can support spatial reasoning and communication in settings that more closely resemble clinical environments.

\subsection{Training and Simulation}
Training and simulation systems remain relatively scarce in AR/MR, which likely reflects the prevalence of VR for immersive rehearsal and fully virtual operating room scenarios. AR/MR nonetheless offers distinct advantages when training involves interaction with physical instruments, anatomical models, or task environments, especially when performance can be assessed through tracking. Existing AR/MR training systems often reuse intraoperative visualization patterns but adapt them to controlled, low-risk settings. Representative approaches include interactive rehearsal using anatomical models for procedural steps such as tumor resection \cite{ghandorh2017development} and AI-based evaluation pipelines that track actions and provide feedback on performance \cite{zhao2020intelligent}.

In this category, the value of AR/MR lies less in supporting conceptual understanding and more in enabling procedural rehearsal under realistic operational constraints. These systems allow users to practice not only task steps, but also tool coordination, line-of-sight management, and interpretation of overlays when alignment is imperfect. Framed this way, anchoring and registration become part of the skill being trained rather than merely technical implementation details. Inter-situ setups can further support progressive learning by enabling view manipulation and error-tolerant guidance before transitioning to fully in-situ workflows.

\subsection{Preoperative Planning}
Preoperative planning tasks leverage AR/MR to help clinicians assess patient-specific anatomy and determine surgical strategies prior to intervention. These systems frequently reconstruct imaging data, most commonly from CT or MRI, into 3D surface models that can be rotated, measured, and annotated to support surgical preparation (\cref{fig:task_taxonomy}). More than half of the systems in this category adopt off-situ spatial integration, showing visualizations in a separate reference frame and not directly aligned with the patient’s body. This setup enables user interaction without real-world registration. It is typically implemented using display technologies such as monitors or tablets \cite{gonzalez2020nextmed,moreta2021combining}, projectors \cite{hansen2010illustrative,schwenderling2022augmented}, or HMDs \cite{cutolo2017new,fischer2020evaluation}.

Although many planning systems are not constrained by strict real-time requirements or direct physical alignment, preoperative planning still depends on an accurate understanding of the spatial relationships among lesions, targets, and surrounding anatomy. The literature often addresses this need through clear, controllable views that support inspection, measurement, and comparison during tasks such as trajectory planning \cite{reitinger2004tools,olexa2024apple} or collaborative surgical discussions \cite{jia2022case}. However, explicit attention to perceptual awareness remains relatively limited in this category. Even if it is typically less critical than in intraoperative guidance, perceptual factors still matter for efficient and reliable planning, particularly when interpreting complex spatial relationships or in-situ planning application.

\subsection{Intraoperative Use}
Typical intraoperative goals include information display \cite{pauly2015machine,katic2013context}, surgical guidance \cite{weidert2019video,tabrizi2015augmented}, and patient positioning \cite{johnson2022patient}. As Fig.~\ref{fig:task_taxonomy} shows, most intraoperative systems anchor imaging \emph{in situ} and use CT imaging modality. They typically present 3D overlays on monitors or OST-HMDs, most often as surface models. This configuration makes perceptual interpretation especially important, because depth ambiguity, occlusion conflicts, and visual clutter may reduce guidance effectiveness once overlays are registered \emph{in situ}. Despite this, only a minority of intraoperative systems report perceptual cue mechanisms or explicitly address depth-related interpretability.


To clarify how intraoperative visualization varies across contexts, we further stratify studies by surgical area. Differences in rigidity, deformation, and physiological motion of the anatomy shape the feasibility of stable registration and the visualization strategies used (\cref{tab:medical_specialty}). The following subsections summarize area-specific patterns and perceptual bottlenecks that emerge under each set of constraints. Although spine-related procedures are often associated with neurosurgery, they also overlap substantially with orthopedic practice. To maintain consistency in our specialty-level comparison, we categorized spine studies under orthopedics.\vspace{5pt}

\noindent \textbf{\underline{Neurosurgery:}}
Multimodal inputs appear more often in neurosurgery, with MRI occurring relatively frequently alongside CT (\cref{fig:area_taxonomy}). This likely reflects the need to understand not only the lesion itself, but also its relationship to nearby critical structures, such as vessels, functional regions, and white matter. 
To support spatial reasoning across these dense anatomical relationships, neurosurgical systems tend to use in-situ 3D overlays. However, the information needs to be presented in such a way that does not increase visual clutter, misleading occlusion, or depth-ordering ambiguity. 
\vspace{3pt}

\noindent \textbf{\underline{Orthopedic:}}
Orthopedic intraoperative systems primarily rely on CT and X-rays. They often use 3D model overlays, but a subset of systems also presents 2D fluoroscopic or X-ray images directly within the augmented view (\cref{fig:area_taxonomy}). This pattern likely reflects the procedural realities of orthopedic surgery, where fluoroscopy imaging is deeply integrated into the surgical workflow
. Because the operative field is dominated by rigid bony anatomy, visualization requirements are often geometry-driven rather than tissue-interpretive. Accordingly, an important design question is not only how to present 3D anatomy, but also how to place and relate 2D intraoperative images to the surgical field in a spatially meaningful way. In this context, AR may reduce reliance on repeated mental reconstruction from conventional fluoroscopic views and better support spatial orientation during orthopedic guidance.\vspace{3pt}

\noindent \textbf{\underline{Head \& Neck:}}
Head \& Neck shows the highest proportion of in-situ use (\cref{fig:area_taxonomy}). This distribution is consistent with the fact that image-to-patient registration is more feasible to achieve. The rigid bony anatomy and accessible facial or dental reference structures facilitate these tasks. At the same time, once registration is achieved, a major remaining challenge becomes the perceptual readability of the overlay at close range, where clutter, occlusion, and depth ordering can quickly reduce interpretability and usability.
\vspace{3pt}

\noindent \textbf{\underline{Thoraco-abdominal:}}
Thoraco-abdominal intraoperative systems use in-situ and off-situ equally often (Fig.~\ref{fig:area_taxonomy}). Although they also rely on the use of CT, ultrasound appears more frequently, consistent with the need for intraoperative updates in deformable and motion-prone environments. While 3D is still common, volume rendering is comparatively more frequent than the intraoperative baseline (\cref{fig:area_taxonomy}). 
One plausible explanation is that the highly deformable nature of thoracoabdominal anatomy increases registration uncertainty, thus favoring strategies that are more tolerant to misalignment. Prior studies have shown a preference for intraoperative cues and volumetric context when surface information alone becomes unreliable due to deformation or partial observability.



\section{Discussion and Future Directions}

\subsection{Visualization Underprioritized in System Design}
\label{sec: discussion-modality}

Despite the inherently visual nature of AR/MR, visualization design remains underemphasized in many medical systems. Across the reviewed literature, visualization choices often appear only loosely tied to modality characteristics, task demands, and perceptual requirements. Many systems rely on generic surface models and standard rendering pipelines, while only a limited subset incorporates perceptual strategies such as contour enhancement, occlusion handling, or focused views, despite their importance for spatial interpretation. 

In neurosurgical visualization, cue selection affects depth judgment and response time \cite{kersten2013evaluation}. In planning environments, display and interaction design influence the mental effort required to transform image information into action \cite{abhari2014training}. Together, these findings suggest that visualization choices should be treated as core design decisions rather than secondary components. Augmented content should provide consistent cues that reduce misleading spatial interpretation, and these choices should be evaluated for their effects on interpretation, task performance, and confidence under realistic workflow constraints.

This underemphasis is also visible over time. While the use of OST-HMDs nearly doubled from 31.0\% (2016–2020) to 59.6\% (2021–2026), the proportion of studies in which perceptual cues could be identified declined from 25.9\% to 8.8\% in the same period. This trend is particularly concerning given that OST-HMDs, while preserving line-of-sight, struggle with occlusion and depth consistency. Results from this search suggest that as medical AR/MR scales toward wider clinical prototyping, explicit perceptual support has failed to keep pace with rapid hardware uptake, potentially leaving critical spatial ambiguities unaddressed. This temporal mismatch should be interpreted cautiously, as it may reflect reporting bias, terminology differences, shifting research priorities, or insufficient methodological and visual detail for unambiguous categorization. Nevertheless, the observed pattern suggests that visualization-specific decisions are often not given the same level of attention as hardware adoption and application expansion. 

The taxonomy introduced in this work provides a foundation for systematically analyzing how visualization choices appear in clinical contexts. While it does not serve as an evaluative framework, it enables reflective analysis of how design decisions relate to imaging modalities, task demands, and user needs.  In Sec.~\ref{sec:design_heuristics} we expand this idea and derive a set of heuristics to support such reflection.

\subsection{Deformable Tissue and Dynamic Anatomy}
One of the most persistent challenges in medical AR/MR visualization is dealing with deformable tissues. Unlike rigid structures such as bones, soft tissues can shift, stretch, or compress during procedures. These changes (often caused by surgical manipulation, physiological motion, or mismatches between preoperative and intraoperative imaging), can fail to reflect the patient’s current anatomy and compromise the accuracy of augmented content. This issue is particularly critical in procedures involving organs such as the liver or lungs, where deformation is substantial and variable.

Recent advances in neural scene representations offer promising directions for addressing this challenge. Methods such as Neural Radiance Fields (NeRF) \cite{mildenhall2021nerf}  and 3D Gaussian Splatting (3DGS) \cite{kerbl20233d} enable continuous, view-consistent reconstructions of anatomy from sparse image inputs, without explicit surface modeling.
Although originally developed for static scenes, recent extensions support real-time rendering \cite{wu20244d} and deformable object modeling \cite{yang2024deformable}. In parallel, these techniques are being explored in medical contexts, including surgical scene reconstruction from endoscopic video \cite{wang2022neural,shu2025seamless}, and direct integration into AR/MR systems \cite{zou2024arthronerf}.
However, applying these methods in clinical AR/MR remains nontrivial. They currently require dense data capture, high computational load, and careful calibration, all of which pose integration barriers in surgical settings. Moreover, their robustness under deformation, occlusion, and varying imaging conditions is still under investigation.
Future work may explore hybrid approaches that integrate them with real-time tracking or biomechanical models to enable adaptive AR/MR visualizations that evolve with patient anatomy during procedures.

\subsection{Context-Aware Design Heuristics}
\label{sec:design_heuristics}
While the taxonomy presented in this work is descriptive rather than evaluative, a qualitative reading of the surveyed literature suggests recurring associations between visualization strategies and clinical contexts. These patterns provide insight into how designers have historically addressed varying clinical demands, but do not indicate which choices are inherently more effective.
Based on these observed patterns, we outline a set of context-aware heuristics as a means to provide design prompts for reflection and alignment.
\vspace{3pt}

\noindent \underline{\textbf{Preserve Modality-specific Strengths}}. As discussed in \cref{sec: discussion-modality}, and shown in \cref{fig:task_taxonomy}, a large portion of the work chose surface rendering as the rendering method. However, a uniform rendering strategy may not be appropriate across all imaging modalities. A practical first step is to evaluate whether the chosen representation compromises the defining imaging characteristics of the modality, like density-based contrast in CT or intensity gradients in MRI. Visualization strategies should then be selected to highlight these characteristics. For example, surface models are well-suited for delineating high-contrast structural boundaries in CT, but should be complemented with volume rendering to capture subcortical or occluded anatomy that may not appear on the surface \cite{kuszyk1996skeletal}. In contrast, MRI emphasizes soft-tissue contrast through smooth intensity variations and subtle anatomical boundaries. Directly converting these signals into rigid surface representations may oversimplify the data and limit the visualization of nuanced tissue differences.
\vspace{3pt}

\noindent \underline{\textbf{Designing Display and Interaction for Clinical Workflow}}. Designers should consider how task timing, sterility, and collaboration shape display and input choices. Head-mounted or spatially aligned in-situ views may suit tasks requiring continuous spatial alignment and hands-free control; handheld or monitor-based views may better support planning, annotation, and shared review. 
\vspace{3pt}

\noindent \underline{\textbf{Fix Ambiguity with Perceptual Cues}}. Accurate depth estimation is a big challenge in AR-based medical visualization (\cref{sec:perceptual}). When depth ambiguity or registration uncertainty could compromise safety or task performance, consider incorporating perceptual supports, such as occlusion management, dynamic visualization, or illustrative cues, to improve spatial interpretability.
\vspace{3pt}

The heuristics presented in this work aim to invite designers to consider how visualization strategies can support the needs and constraints of different clinical scenarios, offering a context-aware scaffold for reflecting on design choices. However, it is important to mention that they are derived from observed practices and not intended as definitive design benchmarks or universal standards.

\section{Conclusion}
Augmented and Mixed Reality offer novel opportunities for visualizing medical imaging and supporting clinical decision-making. In this work, we introduced a taxonomy that serves to classify AR/MR systems used for medical imaging visualization into six dimensions: application, image modality, dimensionality, display technology, visualization anchoring, and perceptual support. Based on a systematic literature review, we examine how these design choices relate to clinical tasks and identify common trends and gaps.

While AR/MR holds strong potential for improving spatial understanding and workflow efficiency, our analysis reveals persistent challenges in modality integration, perceptual design, and adaptability to soft tissue dynamics. Addressing these gaps will require a more visualization-centered development approach and closer collaboration between technical designers and clinical users. 
We expect the taxonomy to support (i) transparent reporting of visualization choices, (ii) cross-context comparison, and (iii) identification of underexplored design combinations, for example using volume rendering more often in oncology workflows where lesion boundaries are unclear and surface models can be misleading.







\acknowledgments{%
  This work was supported by the Graduate Partnership Program (GPP) of the National Institutes of Health (NIH) and the National Institute of Biomedical Imaging and Bioengineering (NIBIB).}%

\bibliographystyle{abbrv-doi-narrow}

\bibliography{_Bibliography}
\end{document}